\documentclass[eqsecnum,showpacs,amsmath,amssymb]{revtex4}
\usepackage{graphics}
\begin{document}
\title{$SU(2)\times SU(2)$ nonlocal quark model with confinement}
\author{A.~E.~Radzhabov}
\author{M.~K.~Volkov}
\affiliation{Bogoliubov Laboratory of Theoretical Physics,
Joint Institute for Nuclear Research, 141980, Dubna, Russia}
\begin{abstract}
The nonlocal version of the $SU(2) \times SU(2)$
symmetric four-quark interaction of the NJL type is considered.
Each of the quark lines contains the form factors. These form
factors remove the ultraviolet divergences in quark loops. The
additional condition on quark mass function $m(p)$ ensures the
absence of the poles in the quark propagator(quark confinement).
The constituent quark mass $m(0)$ is expressed thought the cut-off
parameter $\Lambda$, $m(0)=\Lambda=340$ MeV in the chiral limit.
These parameters are fixed by the experimental value of the weak
pion decay and allow us to describe the mass of the light scalar
meson, strong decay $\rho \rightarrow \pi \pi$ and $D/S$ ratio in
the decay $a_1 \rightarrow \rho \pi$ in satisfactory agreement
with experimental data. 
\end{abstract}
\maketitle
\section{Introduction}
The effective meson Lagrangians obtained on the basis of the local
four-quark interaction of the Nambu--Jona-Lasinio(NJL) type
satisfactorily describe low-energy meson physics
\cite{ZC,Volkov:zb,Vogl:1991qt,Klevansky:qe}.
However, these models contain ultraviolet(UV) divergences and do
not describe quark confinement. Therefore, additional
regularization is necessary in these models. Besides, it is
impossible to take into account the dependence of the amplitude of
different processes on large external momentum in order to
provide quark confinement. This situation restricts the predictive
power of these models. Satisfactory results in these models can
be obtained only for light mesons and interactions at low energies
in the range of $1$ GeV. In order to overcome these restrictions,
it is necessary to consider nonlocal versions of these models which
allow us to remove UV divergences and describe the quark confinement.

A lot of models of this type were proposed in the last few years.
Unfortunately, we cannot give here the full list of references
concerning this activity. Therefore, we will concentrate only on
the direction connected with the nonlocal quark interaction
motivated by the instanton theories
\cite{Shuryak:1981ff,Diakonov:1983hh,Dorokhov:gu,Anikin:rq}.
Recently, a few nonlocal models of the this type were proposed
\cite{Bowler:ir,Plant:1997jr,Dorokhov:2001wx,DRV}. In these models
the nonlocal kernel is taken in the separable form where each
quark line contain form factor following from instanton theories.
These form factors naturally remove UV divergences in quark loops.
Thus, in \cite{Bowler:ir,Plant:1997jr} a nonlocal form factor was
chosen in the Gaussian form $f(p)=\exp(-{p^2}/{\Lambda^2})$ where
$\Lambda$ is the cut-off parameter\footnote{Here and further all
expressions are given in Euclidean domain.}. In
\cite{Dorokhov:2001wx,DRV} it was proposed to use an additional
condition for the form factor $f(p)$ (quark mass function $m(p)$,
respectively) which lead to the absence of the poles in the quark
propagator. Namely, it is supposed that the scalar part of the
quark propagator is expressed through the entire function
\begin{eqnarray}
\frac{m(p)}{p^2+m^2(p)} =
\frac{1}{\mu}\exp\left(-{p^2}/{\Lambda^2}\right),
\end{eqnarray}
where $\mu$ is an additional arbitrary parameter. The similar
condition providing confinement was used in \cite{Efimov,Roberts,Roberts2}.

In this work an analogous condition will be used . However, we
will take into account that each quark line contain square of the form factor
which is expressed through the quark mass
\begin{eqnarray}
\frac{m(p)f^2(p)}{p^2+m^2(p)} \rightarrow
\frac{m^2(p)}{p^2+m^2(p)} = \exp\left(-{p^2}/{\Lambda^2}\right).
\end{eqnarray}
As a result, we obtain a simpler solution for the mass function
than in \cite{Dorokhov:2001wx,DRV}. In our model $m(0)$ and the
cut-off parameter $\Lambda$ have a simple connection in the chiral
limit $m(0)=\Lambda$; the function $m(p)$ contains only one
arbitrary parameter. We fix this parameter by weak pion decay.
Then, for $F_\pi=93$ MeV we have $m(0)=\Lambda=340$ MeV. This
leads to reasonable predictions for the scalar meson mass, width
of the decay $\rho \rightarrow \pi\pi$ and $D/S$ ratio in the
decay $a_1 \rightarrow \rho\pi$, where $D$,$S$ are the partial
waves of this decay .

The paper is organized as follows. In Sect.\ref{mod}, we consider
a nonlocal four-quark interaction and after bosonization derive
the gap equation for dynamical quark mass. The additional
condition for this mass allows us to provide the quark
confinement. In Sect.\ref{pseudoscalar}, the masses and couplings
of the scalar and pseudoscalar mesons are obtained and the main
parameters of the model are fixed. The decay width $\sigma
\rightarrow \pi \pi$ is calculated. In Sect.\ref{axialvector}, the
vector and axial-vector sectors of the model are considered. The
$a_1$ meson mass, decay widths $\rho \rightarrow \pi\pi$, $a_1
\rightarrow \rho\pi$ are calculated. $D/S$ ratio in the decay $a_1
\rightarrow \rho\pi$ is estimated. The $\pi-a_1$ mixing is
studied. The discussion of the obtained results and comparison
with others models is given in the last section.

\section{$SU(2)\times SU(2)$ nonlocal quark interaction\label{mod}}

The $SU(2)\times SU(2)$ symmetric action with the nonlocal four-quark
interaction has the form
\begin{eqnarray}
\mathcal{S}(\bar{q},q)&=&\int d^4x \,\left\{
\bar{q}(x)(i \hat{\partial}_x -m_c)q(x) 
\right.\nonumber \\
&&\quad
+\frac{G_1}{2} \left( J_\pi^a(x) J_\pi^a(x) + J_\sigma(x) J_\sigma(x)\right)
\left.-\frac{G_2}{2} \left( J_\rho^{\mu \, a}(x) J_\rho^{\mu \, a}(x)
+
J_{a_1}^{\mu \, a}(x)J_{a_1}^{\mu \, a}(x)\right) 
\right\}\label{lag},
\end{eqnarray}
where $\bar{q}(x)=(\bar{u}(x),\bar{d}(x))$ are the $u$ and $d$
quark fields, $m_c$ is the diagonal matrix of the current quark
masses, $G_1$ is the coupling constant of the scalar and
pseudoscalar quark currents, $G_2$ is the coupling constant of the
vector and axial-vector quark currents. The nonlocal quark
currents $J_I(x)$ are expressed as
\begin{eqnarray}
J_I(x)=\int
 d^4x_1 d^4x_2 \, f(x_1)f(x_2)\,
\bar{q}(x-x_1) \, \Gamma_I \, q(x+x_2) ,
\label{J}\nonumber
\end{eqnarray}
where $f(x)$ are the nonlocal functions.  
In (\ref{J}) the matrices $\Gamma_I$ are defined as
\begin{eqnarray}
\Gamma_\sigma=\mathbf{1},
\Gamma_\pi^a=i \gamma^5 \tau^a,
\Gamma_\rho^{\mu \, a}=\gamma^\mu \tau^a,
\Gamma_{a_1}^{\mu \, a}=\gamma^5 \gamma^\mu \tau^a, \nonumber
\end{eqnarray}
where $\tau^a$ are the Pauli matrices and $\gamma^\mu,\gamma^5$
are the Dirac matrices.

In this article, we mainly consider the strong interactions. The
electroweak fields may be introduced by gauging the quark field by
the Schwinger phase factors (see, c.f.
\cite{Anikin:rq,Bowler:ir,Plant:1997jr}).

After bosonization the action becomes
\begin{eqnarray}
&&\mathcal{S}(q,\bar{q},\sigma,\pi,\rho,a) =
\int d^4x \left\{
\bar{q}(x)(i \hat{\partial}_x -m_c)q(x) - \frac{1}{2 G_1}\left( \pi^a(x)^2 +\tilde{\sigma}(x)^2\right) 
+ \frac{1}{2 G_2}\left((\rho^{\mu \, a}(x))^2 + (a_1^{\mu \, a}(x))^2\right)
\right.
\nonumber\\
&&\,
+
\left.
\int d^4x_1 d^4x_2 \, f(x-x_1)f(x_2-x)
\bar{q}(x_1)\left( \pi^a(x)i \gamma^5 \tau^a
+\tilde{\sigma}(x) +\rho^{\mu \, a}(x)\gamma^\mu \tau^a
+a_1^{\mu \, a}(x) \gamma^5 \gamma^\mu\tau^a \right)q(x_2) \right\},\label{acti}
\end{eqnarray}
where $\tilde{\sigma},\pi,\rho,a$ are the $\sigma$, $\pi$, $\rho$,
$a_1$ meson fields, respectively. The field $\tilde{\sigma}$ has a
nonzero vacuum expectation value
$<\tilde{\sigma}>_0=\sigma_0\neq0$. In order to obtain a physical
scalar field with zero vacuum expectation value, it is necessary to
shift the scalar field as $\tilde{\sigma}=\sigma+\sigma_0$. This
leads to the appearance of the quark mass function $m(p)$ instead of
the current quark mass $m_c$
\begin{eqnarray}
m(p)=m_c+m_{dyn}(p),\label{gap1}
\end{eqnarray}
where $m_{dyn}(p)=-\sigma_0f^2(p)$ is the dynamical quark mass. From
the action, eq. (\ref{acti}), by using
\begin{eqnarray}
\left\langle \frac{\delta S}{\delta \sigma}\right\rangle_0 = 0,\nonumber
\end{eqnarray}
one can obtain the gap equation for the dynamical quark mass
\begin{eqnarray}
m_{dyn}(p)= G_1 \frac{8  N_c }{(2 \pi)^4}f^2(p) \int d^4_Ek
f^2(k) \frac{m(k)}{k^2+m^2(k)}.\label{gap}
\end{eqnarray}
The right-hand side of this equation is the tadpole of the quark
propagator taken in the Euclidean domain.
Equations (\ref{gap1}),(\ref{gap}) have the following solution:
\begin{eqnarray}
m(p)=m_c+(m_q-m_c)f^2(p),\label{gap2}
\end{eqnarray}
where $m_q = m(0)$.

In order to provide quark confinement we propose the following
anzatz for the quark mass function $m(p)$. We suppose that mass
satisfies the following condition in the chiral limit
\begin{eqnarray}
\frac{m^2(p)}{m^2(p)+p^2} = \exp
\left(-{p^2}/{\Lambda^2}\right).\label{neqq1}
\end{eqnarray}
The form of the left-hand side of this equation coincides with
the integrand in the gap equation (\ref{gap}).
From eq.(\ref{neqq1}) we obtain the following solution
\footnote{Here only the positive solution will be used.}:
\begin{eqnarray}
m(p)=\left(\frac{p^2}{\exp
\left({p^2}/{\Lambda^2}\right)-1}\right)^{1/2} ;\label{neqq3}
\end{eqnarray}
here we have only one free parameter $\Lambda$; $m(p)$ does not
have any singularities in the whole real axis and exponentially
drops as $p^2 \rightarrow \infty$ in the Euclidean domain. From
eq.(\ref{gap2}) follows that  the form factors have a similar
behavior that provides the absence of UV divergences in our model.
At $p^2=0$ the mass function is equal to the cut-off parameter
$\Lambda$, $m(0)=\Lambda$. The pole part of the quark propagator
also does not contain singularities that provide quark confinement
\footnote{Note that similar functions were used in
\cite{Roberts,Roberts2,EfimovC} in order to describe the quark
confinement. }
\begin{eqnarray}
\frac{1}{m^2(p)+p^2}=\frac{1-\exp
\left(-{p^2}/{\Lambda^2}\right)}{p^2}.
\end{eqnarray}

When taking into account the current quark mass
eq.(\ref{neqq1}) can be modified as follows:
\begin{eqnarray}
\frac{m^2(p)-m_c^2}{m^2(p)+p^2}= \exp
\left(-{\left(p^2+m_c^2\right)}/{\Lambda^2}\right)\label{ne1}.
\end{eqnarray}
Here $m_c^2$ is introduced in the form that conserves the
analytical properties of the mass function $m(p)$. Then the mass
function takes the form
\begin{eqnarray}
m(p)=\left(\frac{m_c^2+p^2 \exp
\left(-{\left(p^2+m_c^2\right)}/{\Lambda^2}\right)}{1-\exp
\left(-{\left(p^2+m_c^2\right)}/{\Lambda^2}\right)}\right)^{1/2}
\end{eqnarray}

\section{Pseudoscalar and scalar mesons\label{pseudoscalar}}

Let us consider the scalar and pseudoscalar mesons. The meson
propagators are given by
\begin{eqnarray}
D_{\sigma,\pi}(p^2)=\frac{1}{-G_1^{-1}+\Pi_{\sigma,\pi}(p^2)}=
\frac{g^{2}_{\sigma,\pi}(p^2)}{p^2-M_{\sigma,\pi}^2}
\label{mesonprop},
\end{eqnarray}
where $M_{\sigma,\pi}$ are the meson masses, $g_{\sigma,\pi}(p^2)$
are the functions describing renormalization of the meson fields
and $\Pi_{\sigma,\pi}(p^2)$ are the polarization operators defined by
\begin{eqnarray}
\Pi_{\sigma,\pi}(p^2)=
i \frac{2 N_c}{(2\pi)^4} \int d^4k f^2(k_-^2)f^2(k_+^2)
\mathrm{Sp}\left[ \mathrm{S}(k_-)\Gamma_{\sigma,\pi} \mathrm{S}(k_+) \Gamma_{\sigma,\pi}
\right], \label{poloper}
\end{eqnarray}
where $k_\pm=k \pm p/2$.

For calculation of these integrals it is necessary to rewrite
these expressions in the Euclidean space where the
form-factors(and quark masses) are the exponentially decreasing
functions. Then eq.(\ref{poloper}) takes the form:
\begin{eqnarray}
\Pi_{\sigma,\pi}(p^2)=
\frac{2 N_c}{(2\pi)^4 m_q^2} \int d_E^4k 
\frac{P_{\sigma,\pi}(k^2,p^2,p \cdot
k)}{(k_+^2+m(k_+^2)^2)(k_-^2+m(k_-^2)^2)} \label{poloperE}
\end{eqnarray}
The functions $P_{\sigma,\pi}(k^2,p^2,p \cdot k)$ are the Dirac
trace multiplied by $m(k_+)$,$m(k_-)$. In eq.(\ref{poloperE}) all
momenta are Euclidean. In the description of the meson properties
it is necessary to make the analytical continuation of this
expression over external momenta $p$ to the Minkowski space. Let
us emphasize that at our anzatz for a quark mass function only
the functions $P_{\sigma,\pi}(k^2,p^2,p \cdot k)$ contains
nonanalytical root cuts, whereas there is no problems with
analytical continuation of the denominator.

The meson masses $M_{\sigma,\pi}$ are found from the position
of the pole in the meson propagator
\begin{eqnarray}
\Pi_{\sigma,\pi} (M_{\sigma,\pi}^2) = G_1^{-1},\label{pii}
\end{eqnarray}
and the constants
$g_{\sigma,\pi}(M_{\sigma,\pi}^2)$ are given by
\begin{eqnarray}
g_{\sigma,\pi}^{-2}(M_{\sigma,\pi}^2) = \frac{d \Pi_{\sigma,\pi} (p^2)}{d
p^2}|_{p^2=M_{\sigma,\pi}^2}.\label{gm}
\end{eqnarray}

Firstly, let us consider this model in the chiral limit. The pion
constant $g_\pi(0)$ is not depend on parameter $\Lambda$ and takes
the form
\begin{eqnarray}
g_\pi^{-2}(0) = \frac{N_c}{4
\pi^2}\left(\frac{3}{8}+\frac{\zeta(3)}{2} \right) \, , \,
g_\pi(0) \approx 3.7,\label{gpi}
\end{eqnarray}
here $\zeta$ is the Riemann zeta function.

The gap equation has the simple form
\begin{eqnarray}
G_1\Lambda^2=\frac{2\pi^2}{N_c}. \label{G}
\end{eqnarray}
The quark condensate is
\begin{eqnarray}
\langle \bar{q}q \rangle_0=-\frac{N_c}{4 \pi^2}\int
\limits_0^\infty du \, u \frac{m(u)}{u+m^2(u)}. \label{cond}
\end{eqnarray}
The Goldberger-Treiman relation is fulfilled in the model of this
kind \cite{Bowler:ir,Plant:1997jr,Anikin:rq,DRV}
\begin{eqnarray}
F_\pi=\frac{m_q}{g_\pi}. \label{ffpi}
\end{eqnarray}
From eqs.(\ref{gpi}),(\ref{ffpi}) the value $\Lambda=m_q=340$ MeV
is obtained for $F_\pi=93$ MeV. Then, from eqs.(\ref{G}),
(\ref{cond}) we obtain
\begin{eqnarray}
G_1=56.6 \, \mathrm{GeV} , \quad
\langle \bar{q}q \rangle_0=-(188 \, \mathrm{MeV})^3 \label{par1}.
\end{eqnarray}

In the description of pion mass it is necessary to introduce the
nonzero current quark mass $m_c$. In our model $M_\pi^{2} \ll
\Lambda^2$. Therefore, we can consider only the lowest order of the
expansion in small $p^2$. Then, one gets from eq.(\ref{mesonprop})
\begin{eqnarray}
M_\pi^{2} = g_\pi^2(0) \left( G_1^{-1}-\frac{N_c}{2 \pi^2}\int
\limits_0^\infty du \, u \frac{f(u)^4}{u+m^2(u)}\right).
\end{eqnarray}
By using the expression for $G_1$ from the gap equation
(\ref{gap}), the Gell-Mann--Oakes--Renner relation can be
reproduced
\begin{eqnarray}
M_\pi^{2} &=&
 -2 \frac{m_c \langle \bar{q}q \rangle_0}{F_\pi^2}+ O(m_c^2) . \label{mpi}
\end{eqnarray}
From eq.(\ref{mpi}) with $M_\pi=140$ MeV we can estimate the value
of the current quark mass $m_c \approx 13$ MeV. Other model
parameters in this case change very little
\begin{eqnarray}
\Lambda=343 \, \mathrm{MeV}, \quad g_\pi(M_\pi)=3.57 ,
\quad
 G_1=56.5 \mathrm{GeV} ,\, 
 \quad \langle \bar{q}q \rangle_0=-(189 \, \mathrm{MeV})^3 \label{par2}.
\end{eqnarray}
Therefore, in calculations of the amplitudes of various processes
we can use the values of parameters taken in the chiral limit.

With the help of the parameters (\ref{par1}) we get for sigma
meson $M_\sigma=420$ MeV and $g_\sigma(M_\sigma)=3.85$. The
amplitude of the decay $\sigma \rightarrow \pi\pi$ is equal to
$A_{(\sigma\rightarrow\pi^+\pi^-)}=1.67$ GeV. Then, the total
decay width is $\Gamma_{(\sigma \rightarrow\pi\pi)} = 150$ MeV.
Comparing these results with experimental data one finds that
$M_\sigma$ is in satisfactory agreement with experiment
$M^{\mathrm{exp}}_\sigma=400-1200$; however, the decay width is very small
$\Gamma^{\mathrm{exp}}_\sigma=600-1000$.

\section{Vector and axial-vector mesons\label{axialvector}}

The propagators of the vector and axial-vector mesons have the
transversal and longitudinal parts
\begin{eqnarray}
D^{\mu \nu}_{\rho,a_1}=T^{\mu \nu}D^{T}_{\rho,a_1}+L^{\mu \nu}D^{L}_{\rho,a_1} \label{m2prop},
\end{eqnarray}
where $T^{\mu \nu}=g^{\mu \nu}-{p^\mu p^\nu}/{p^2}\,,\,L^{\mu \nu}={p^\mu p^\nu}/{p^2}$ and
\begin{eqnarray}
D^{T}_{\rho,a_1} =\frac{1}{G_2^{-1}+\Pi^{T}_{\rho,a_1} (p^2)}=
\frac{g^{2}_{\rho,a_1}(p^2)}{M_{\rho,a_1}^2-p^2} ,\quad 
D^{L}_{\rho,a_1} =\frac{1}{G_2^{-1}+\Pi^{L}_{\rho,a_1} (p^2)}
.\label{grho}
\end{eqnarray}
Here, $\Pi^{T}_{\rho,a_1}$ and $\Pi^{L}_{\rho,a_1}$ are the transversal and
longitudinal parts of the polarization operator $\Pi^{\mu\nu}_{\rho,a_1}(p^2)$
\begin{eqnarray}
\Pi_{\rho,a_1}^{\mu\nu}(p^2)=
 i \frac{2 N_c}{(2\pi)^4} \int d^4k f^2(k_-)f^2(k_+)
\mathrm{Sp}\left[ \mathrm{S}(k_-)\Gamma_{\rho,a_1} \mathrm{S}(k_+) \Gamma_{\rho,a_1}
\right].\nonumber
\end{eqnarray}
The constant $G_2$ is fixed by the $\rho$-meson mass
\begin{eqnarray}
G_2^{-1} = -\Pi^{T}_{\rho}(M_{\rho})\nonumber
\end{eqnarray}
and $G_2=6.5$ GeV$^{-2}$. Then the $a_1$-meson mass is equal to
$970$ MeV.

The constants $g_{\rho,a_1}(M_{\rho,a_1}^2)$ are equal to
\begin{eqnarray}
g^{-2}_{\rho,a_1}(M_{\rho,a_1}^2)=-\frac{d \Pi^{T}_{\rho,a_1} (p^2)}{d
p^2}|_{p^2=M_{\rho,a_1}^2}.\label{gm2}
\end{eqnarray}
From eq.(\ref{gm2}) we obtain $g_\rho(M_\rho)=1.23$,
$g_a(M_{a_1})=1.33$. At $p^2=0$ we have $g_\rho(0)\approx 2$,
$g_a(0)\approx 2.5$. (see also Fig.\ref{g}).

The decay $\rho \rightarrow \pi\pi$ is described by the triangle
quark diagram. The amplitude for the process is
\begin{eqnarray}
A^\mu_{(\rho \rightarrow \pi \pi)}=a_{(\rho \rightarrow \pi\pi)} (q_1-q_2)^\mu
\end{eqnarray}
where $q_i$ are momenta of the pions. We obtain $a_{(\rho
\rightarrow \pi\pi)} = 5.72$ and the decay width
$\Gamma_{(\rho\rightarrow \pi \pi)}= 135 \, \mathrm{MeV}$ which is
in qualitative agreement with the experimental value $149.2 \pm
0.7$ MeV \cite{Hagiwara:fs}.

The decay $a_1 \rightarrow \rho\pi$ is described in a similar
manner. The amplitude for the process $a_1 \rightarrow \rho\pi$ is
\begin{eqnarray}
A^{\mu\nu}_{(a_1 \rightarrow \rho \pi)}=
a_{(a_1 \rightarrow \rho \pi)} g^{\mu \nu}
+
b_{(a_1 \rightarrow \rho \pi)} p^{\nu} q^{\mu}
\end{eqnarray}
where $p,q$ are momenta of $a_1$,$\rho$ mesons, respectively. We
obtain $a_{(a_1 \rightarrow \rho \pi)} = 2.68$ GeV, $b_{(a_1
\rightarrow \rho \pi)} = 16.71$ GeV$^{-1}$. Amplitude of the decay
$a_1 \rightarrow \rho \pi$ contains D and S waves. The ratio of
these waves has the form(see \cite{Plant:1997jr,Roberts2}):
\begin{eqnarray}
&&D/S=-\sqrt{2} \frac{(E_\rho-M_\rho)a_{(a_1 \rightarrow \rho
\pi)}+b_{(a_1 \rightarrow \rho \pi)} M_{a_1}
|\overrightarrow{q}|^2} {(E_\rho+2M_\rho)a_{(a_1\rightarrow \rho
\pi)}
+b_{(a_1 \rightarrow \rho \pi)}M_{a_1} |\overrightarrow{q}|^2}=
-0.06,\\
&&|\overrightarrow{q}|^2=\lambda
(M_{a_1}^2,M_{\rho}^2,M_{\pi}^2)/(2
M_{a_1})^2, \, E_\rho^2=M_\rho^2+|\overrightarrow{q}|^2, \, 
\lambda (a,b,c)=a^2+b^2+c^2-2ab-2ac-2bc\nonumber
\end{eqnarray}
This ratio is in
satisfactory agreement with experimental data $D/S^{\mathrm{exp}}=
-0.108\pm0.016$. The decay width equals $\Gamma_{(a_1 \rightarrow
\rho \pi)}=90$ MeV. This value is noticeably smaller than
experiment $250-600$ MeV \cite{Hagiwara:fs}.

The longitudinal component of the $a_1$-meson field is mixed with
the pion. The amplitude describing this mixing has the form
\begin{eqnarray}
A^\mu_{(\pi\rightarrow a_1)}= i C_{(\pi\rightarrow a_1)}(p^2)
p^\mu ,\nonumber
\end{eqnarray}
where $p$ is the momentum of the pion, and $C_{(\pi\rightarrow
a_1)}(0)$ in the chiral limit is equal to $190$ MeV. The additional
pion kinetic term from the $\pi-a_1$ mixing is $\Delta L_{kin} =
\Delta \cdot {p^2} \pi^a(p)^2 /2 $. In the chiral limit $\Delta$
is equal to
\begin{eqnarray}
\Delta &=&\frac{C_{(\pi\rightarrow a_1)}^2(0)}
{g^2_{a_1}(0)(G_2^{-1}+\Pi^{L}_{a_1}(0))} \approx
C_{(\pi\rightarrow a_1)}^2(0) G_2/g^2_{a_1}(0)
\approx 0.04.
\end{eqnarray}
$\Delta$ is small; therefore, the effect of the $\pi-a_1$ mixing
can be neglected.

\section{Discussion and conclusion}

In this work we have considered a possibility of constructing the
$SU(2) \times SU(2)$ symmetric nonlocal chiral quark model
providing the absence of UV divergences and quark confinement.
These features of the model are specified by the nonlocal kernel
which appears in the four-quark interaction. Such a structure of
the four-quark interaction can be motivated by the instanton
interactions \cite{Diakonov:1983hh,Dorokhov:gu,Anikin:rq}.

\begin{figure}
\hspace{0.3cm}
\resizebox{0.30\textwidth}{!}{\includegraphics{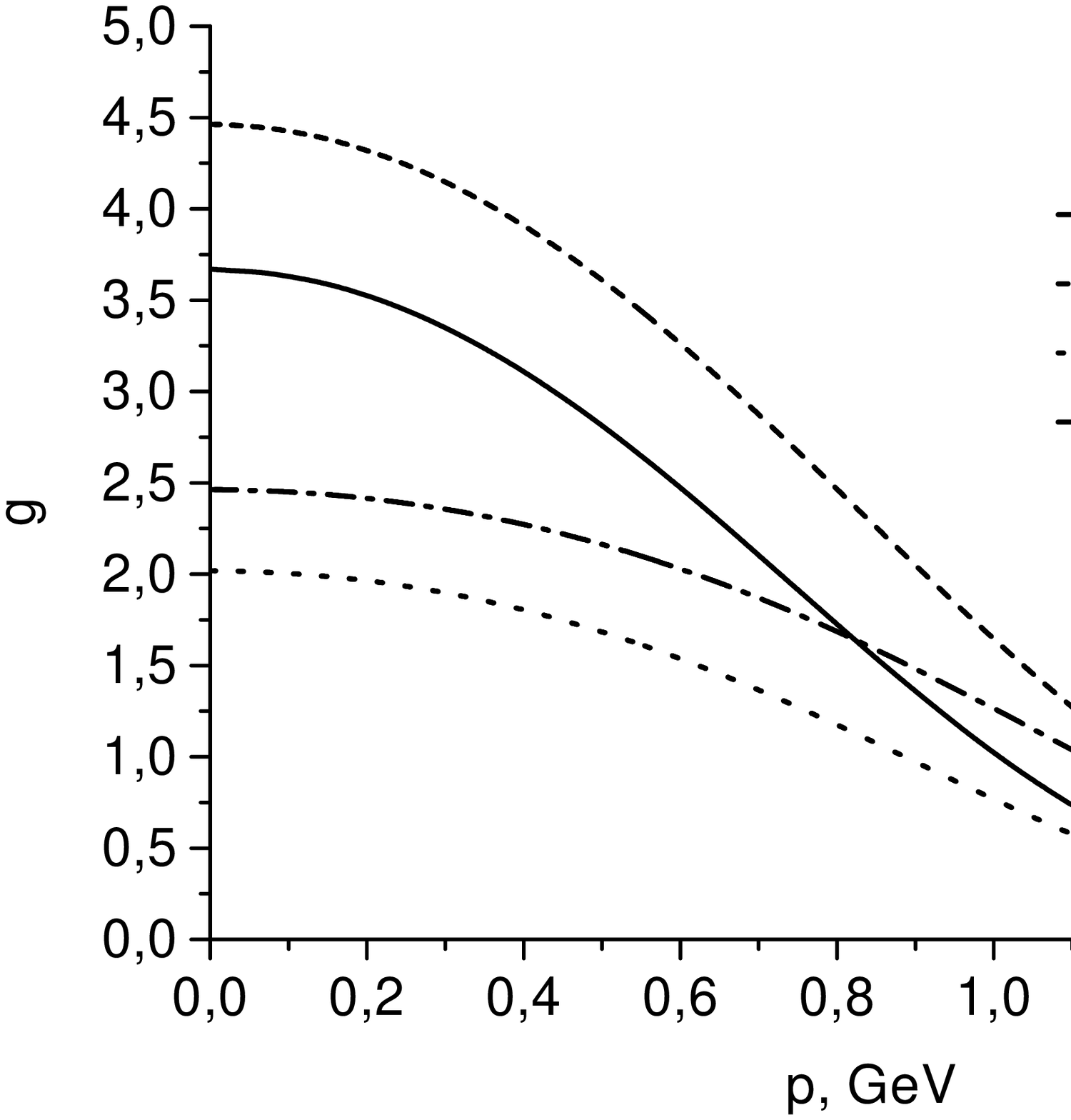}}
\caption{Momentum dependence of the mesons renormalization
functions.} \label{g}
\end{figure}

The pseudoscalar, scalar, vector and axial-vector mesons have been
considered in the framework of this model. The masses and strong
coupling constants of the mesons were described. It was shown that
the functions describing renormalization of the meson fields
noticeably decreased at large $p^2$ in the physical domain(see
Fig.\ref{g}).

Among satisfactory predictions of the model are the mass of the
$\sigma$-meson , the decay width $\rho \rightarrow \pi \pi$ and
the $D/S$ ratio in the decay $a_1 \rightarrow \rho \pi$.

However, in the description of the $a_1$ meson mass and decay
widths $\sigma \rightarrow \pi\pi$ , $a_1 \rightarrow \rho \pi$
our results are noticeably smaller than experimental data. Note
that the width of the decay $a_1 \rightarrow \rho \pi$ strongly
depends on mass of the $a_1$ meson. Indeed, for $M_{a_1}=1.26$ GeV
we have $\Gamma_{(a_1 \rightarrow \rho \pi)}\approx 200$ MeV that
is in qualitative agreement with experiment.

It is useful to compare the obtained results with the analogous
results obtained in the local NJL model \cite{Volkov:zb} and other
nonlocal models with quark interaction of separable type
\cite{Plant:1997jr,Roberts2}.

Remind that in the local NJL model the cut-off parameter
$\Lambda^{(NJL)}=1.2$ GeV and the constituent quark mass $m=280$
MeV are used. These parameters are fixed by the decays $\pi
\rightarrow \mu \nu$($f_\pi=93$ MeV) and $\rho \rightarrow \pi
\pi$($g_\rho=6.14$). These parameters lead to the quark condensate
$\langle\bar{q}q \rangle_0=-(293 \, \mathrm{MeV})^3$ and current
quark mass $m_c=3$ MeV. In the present model $m_0=340$ MeV plays
the role of the constituent quark mass, whereas our parameter
$\Lambda=340$ MeV corresponds to the effective cut-off parameter
$\Lambda^{\mathrm{eff}}\approx 800$ MeV. As a result, we obtain
$\langle \bar{q}q \rangle_0=-(188 \, \mathrm{MeV})^3$ and
$m_c=13$. Remind that these values correspond to the physical pion
mass.

Let us consider $\pi - a_1$ mixing in these models. In the local
NJL model the amplitude describing the $\pi-a_1$ mixing equals
$A^{\mu \, (NJL)}_{(\pi \rightarrow  a_1)} = i \sqrt{6} m p^\mu$.
Therefore, the coefficient $C_{(\pi \rightarrow  a_1)}^{(NJL)}$ in
the equals $680$ MeV. This value is $3.5$ times larger than in the
present model. As a result, it leads to the noticeable additional
renormalization of the pion field in the local NJL model
$\tilde{g}_\pi^{(NJL)}=g_\pi^{(NJL)}\cdot Z^{1/2}=m/f_\pi$, where
$Z=\left(1-{6m^2}/{M_{a_1}^2}\right)^{-1}\approx 1.4$ and
$g_\pi^{(NJL)}=g_\sigma^{(NJL)}$, in our model $Z=1.04$.
Therefore, in the local NJL model the $\pi-a_1$ mixing plays a
more important role.

Let us compare also the amplitude of the decay width $\sigma
\rightarrow \pi\pi$ in these models. In the local NJL model
without taking into account the $\pi - a_1$ mixing in the external
pion legs this amplitude equals $A^{(NJL)}_{\sigma \rightarrow
\pi^+\pi^-}=4 m \tilde{g}_\pi^{(NJL)} Z^{1/2}=4$ GeV. This
amplitude is $2.4$ times larger than in the present model.
However, after taking into account the $\pi-a_1$ mixing this amplitude
takes the form $A^{(NJL)}_{\sigma \rightarrow \pi^+ \pi^-}=4 m
\tilde{g}_\pi^{(NJL)} Z^{-3/2}=2$ GeV. This is close to the
amplitude obtained in our work. This leads to a noticeable decrease
in the decay width $\Gamma_{\sigma \rightarrow \pi \pi}=190$ MeV
which also becomes smaller than experimental data and is in
qualitative agreement with the result of our model. The mass of
the $\sigma$ meson in NJL is $570$ MeV and approximately 30 \%
larger than the
result obtained here. Both the values do not contradict the
experimental data.

The decay $\rho \rightarrow \pi\pi$ in \cite{Volkov:zb} is
used for fitting model parameters while in our model we predict
it. The mass of the $a_1$ meson obtained in our model practically
coincides with the mass predicted in the local NJL model,
$M_{a_1}^{(NJL)} \approx 1$ GeV.

In what follows we would like to compare our result with the
nonlocal model \cite{Plant:1997jr}. In this model a similar
separable instanton-motivated form of the interaction is also
used. The main difference of our model with that of
\cite{Plant:1997jr} is connected with an additional requirement on
a quark propagator providing quark confinement. The quark mass
function in our model contains only one arbitrary parameter
instead of two parameters in \cite{Plant:1997jr}. In spite of less
freedom in choosing model parameters, our results is close to the
results obtained in \cite{Plant:1997jr}(see table 1).

It is interesting also to compare our results with those obtained
in \cite{Roberts2}, where the quark propagator is expressed
through the entire functions which are similar to the function
used in our work (see eq.(\ref{neqq1})). The quark interaction in
this work has a separable type which is obtained from the
quark-gluon interaction with the help of a modified gluon
propagator. In this work, the decays $\rho \rightarrow \pi \pi$
and $a_1 \rightarrow \rho \pi$ also have been calculated. The
decay ratio $D/S$ is close to ours while the decay widths strongly
differs (see table 1).

The failure of the local NJL model and its nonlocal extensions
to describe the $\sigma$-meson is expectable.
Similar problems appeared in the QCD sum rule
method. In the scalar channel with vacuum quantum numbers the
corrections from different sources may be valuable. Indeed, it has
recently been shown that the $1/N_c$ corrections in this channel
are rather big\cite{Plant:2000ty}, and the Hartree - Fock
approximation may be inadequate in this case. Moreover, for a
correct description of the scalar meson it is necessary to take
into account the mixing with the four-quark state
\cite{Jaffe:1976ig} and the scalar glueball \cite{Volkov:ns}.

In future, we plan to describe electromagnetic interactions in
the framework of this model, calculate the e.m. pion radius,
polarizability of the pion and consider the processes $\pi^0
\rightarrow \gamma \gamma$, $\gamma^\ast \rightarrow \gamma\pi$ in
a wide domain of photon virtuality.
We also plan to generalize this model to the $U(3) \times U(3)$ chiral group by introducing new
parameters: mass of the strange quark $m_s$ and the cut-off $\Lambda_s$
which allows us to describe intrinsic properties and interactions
of strange mesons.

The authors thank A. E. Dorokhov for fruitful collaboration, and
D. Blaschke, C. D. Roberts and V. L. Yudichev for useful discussions.
The work is supported by RFBR Grant no. 02-02-16194 and the
Heisenberg--Landau program.

\clearpage

\begin{center}
\begin{tabular}{| c | c | c | c | c | c | c | c |}
\hline
Quantity & Our model & \cite{Volkov:zb}  & \multicolumn{2}{|c|}{\cite{Plant:1997jr}} & \multicolumn{2}{|c|}{\cite{Roberts2}} & \cite{Hagiwara:fs} \\
\hline
$M_{\sigma}$ (MeV) & $420$ & $570$ & $443.2$ & $465.8$ & & & $400-1200$ \\
$\Gamma_{\sigma \rightarrow \pi \pi}$ (MeV) & $150$ & $190$ & $108$ &$135.1$ & & & $600-1000$ \\
$\Gamma_{\rho \rightarrow \pi \pi}$  (MeV) & $135$ & $150$ & $126$ & $114$ & $356$ & $259$ & $149.2\pm0.7$ \\
$M_{a_1}$  (MeV) & $970$ & $1030$ & $946.8$ & $1061.5$ & \multicolumn{2}{|c|}{$1340$} & $1230\pm40$ \\
$\Gamma_{a_1 \rightarrow \rho \pi}$ (MeV) & $90$ & $290$ & $44$ & $376.2$ & 4020 & $385$ & $250-600$ \\
$D/S$ & $-0.06$ & & $-0.048$ & $-0.087$ & $-0.092$ & $-0.075$ & $-0.108\pm0.016$ \\
\hline
\end{tabular}
\end{center}
\noindent Table 1. The comparison of the physical results are
obtained in local and nonlocal quark models. In local NJL model
the decay width $\rho \rightarrow \pi\pi$ is used for fitting of
the model parameters. Two set of values corresponding to the
different choice of model parameters are given in column 4, 5 (see
table 4 in \cite{Plant:1997jr} and table I in \cite{Roberts2}).

\end{document}